\documentclass{aastex631}

\usepackage[T1]{fontenc}

\usepackage{multirow}
\usepackage{amsmath}
\usepackage{hhline}
\usepackage{array, makecell}
\usepackage{booktabs, tabularx}
\usepackage{comment}

\newcommand{\gray}{$\gamma$-ray}
\newcommand{\grays}{$\gamma$-rays}

\usepackage{multirow}

\usepackage{xcolor}
\newcounter{tr}
\ifnum \value{tr}>5
\newcommand{\deletedD}[1]{{\color{red} Damien - Deleted: } \sout{#1}}

\newcommand{\authorcommentD}[1]{{\color{purple} Damien - Comment :} {\color{cyan} #1}}

\else
\newcommand{\deletedD}[1]{}

\newcommand{\authorcommentD}[1]{}

\fi
\begin{document}

\title{Modeling blazar broadband emission with convolutional neural networks  - II. External Compton model}
%\title[A Soprano Table model]{A Soprano Table model}

\shorttitle{Convolutional neural network for external Compton model}

\shortauthors{Sahakyan et al.}

\author[0000-0003-2011-2731]{N. Sahakyan}
\affiliation{ICRANet-Armenia, Marshall Baghramian Avenue 24a, Yerevan 0019, Armenia}

\author[0000-0003-4477-1846]{D. B\'egu\'e}
\affiliation{Bar Ilan University, Ramat Gan, Israel}

\author[0009-0007-4522-5501]{ A. Casotto}
\affiliation{Chief Scientist, Altair, 100 Mathilda Place, Suite 650 Sunnyvale, CA 94086}

\author[0000-0002-8852-7530]{H. Dereli-B\'egu\'e}
\affiliation{Bar Ilan University, Ramat Gan, Israel}

\author[0000-0002-2265-5003]{P. Giommi}
\affiliation{Associated to INAF, Osservatorio Astronomico di Brera, via Brera, 28, I-20121 Milano, Italy}
\affiliation{Center for Astrophysics and Space Science (CASS), New York University Abu Dhabi, PO Box 129188 Abu Dhabi, United Arab Emirates}
\affiliation{Institute for Advanced Study, Technische Universit{\"a}t M{\"u}nchen, Lichtenbergstrasse 2a, D-85748 Garching bei M\"unchen, Germany}

\author[0000-0002-0031-7759]{S. Gasparyan}
\affiliation{ICRANet-Armenia, Marshall Baghramian Avenue 24a, Yerevan 0019, Armenia}

\author[0000-0002-3777-7580]{V. Vardanyan}
\affiliation{ICRANet-Armenia, Marshall Baghramian Avenue 24a, Yerevan 0019, Armenia}

\author[0009-0007-7798-2072]{M. Khachatryan}
\affiliation{ICRANet-Armenia, Marshall Baghramian Avenue 24a, Yerevan 0019, Armenia}

\author[0000-0001-8667-0889]{A. Pe{'}er}
\affiliation{Bar Ilan University, Ramat Gan, Israel}

\begin{abstract}
In the context of modeling spectral energy distributions (SEDs) for blazars, we extend the method that uses a convolutional neural network (CNN) to include external inverse Compton processes. The model assumes that relativistic electrons within the emitting region can interact and up-scatter
external photon originating
from the accretion disk, the broad-line region, and the torus, to produce the observed
high-energy emission. We trained the CNN on a numerical model that accounts for
the injection of electrons, their self-consistent cooling, and pair creation-annihilation
processes, considering both internal and all external photon fields. 
Despite the larger number of parameters compared to the synchrotron self-Compton
model and the greater diversity in spectral shapes, the CNN enables an
accurate computation of the SED for a specified
set of parameters. The performance of the CNN is demonstrated by fitting the SED of two flat-spectrum radio quasars, namely
3C 454.3 and CTA 102, and obtaining their parameter posterior distributions. For the
first source, the available data in the low-energy band allowed us to constrain the
minimum Lorentz factor of the electrons, \(\gamma_{\rm min}\), while for the second source,
due to the lack of these data, \(\gamma_{\rm min} = 10^2\) was set. We used the obtained parameters to investigate the energetics of the system. The model
developed here, along with one from \citet{BSD23}, enables self-consistent,
in-depth modeling of blazar broadband emissions within leptonic scenario.
\end{abstract}


%BL Lacertae objects: general -- radiation mechanisms: non-thermal -- methods: numerical}

\section{Introduction}
Blazars are a sub-type of active galactic nuclei characterized by jets that are
closely aligned with the observer's line of sight \citep{1995PASP..107..803U}.
Their emissions, which are Doppler-enhanced and highly variable, span an extensive
range of frequencies from the radio to the high
energy (HE; $E>100$ MeV) and very high energy (VHE; $E>100$ GeV) \gray\ bands
\citep{2017A&ARv..25....2P}. This wide spectral range makes blazars prime candidates
for multiwavelength studies. Blazars are typically classified into two major
categories based on the prominence of emission lines in their optical spectra:
BL Lacertae objects (BL Lacs), which exhibit weak or absent emission lines, and
Flat Spectrum Radio Quasars (FSRQs), known for their prominent optical emission
lines. However, this classification extends beyond the mere presence or absence
of emission lines: it also reflects differences in their physical
properties, including the origin of their emissions.

The spectral energy distribution (SED) of blazars
is typically characterized by a double-humped structure. The first hump, peaking
in the infrared (IR) through X-ray bands, is attributed to synchrotron emission
from relativistic electrons within the jet. The second hump, peaking in the \gray\ band,
can be explained by two different mechanisms based on the particle at the origin of the
emission. First, in the leptonic scenario considered here, the HE emission arises from inverse
Compton scattering of either low-energy synchrotron photons \citep[Synchrotron
Self-Compton model (SSC), see \textit{e.g.}][]{1985A&A...146..204G, 1992ApJ...397L...5M,
1996ApJ...461..657B}, or external thermal photons, resulting in the
so-called External Inverse Compton (EIC) model. These
photons can originate from the accretion disk, the broad-line region (BLR), or the
dusty torus \citep[see \textit{e.g.}][]{1994ApJ...421..153S, 1992A&A...256L..27D,
1994ApJS...90..945D, 2000ApJ...545..107B}. SSC models are generally effective in
explaining the broadband data from BL Lacs, while EIC models are usually favored
for FSRQs. Indeed for the latter class, the peak luminosity
of the HE hump strongly dominates over that of the low energy hump.
In the leptonic model considered in this paper, this is referred to as Compton dominance.

The second scenario is the hadronic model, where HE/VHE emission is either from direct
proton synchrotron emission \citep{2001APh....15..121M} or from emissions by secondaries
produced in photo-pion and photo-pair
interactions \citep{1993A&A...269...67M, 1989A&A...221..211M,
2001APh....15..121M, mucke2, 2013ApJ...768...54B, 2015MNRAS.447...36P, GBS22}.
The potential detection of VHE neutrinos from the direction of blazars, such as
TXS 0506+056 \citep{2018Sci...361..147I, 2018Sci...361.1378I, 2018MNRAS.480..192P}
and PKS 0735+178 \citep{2023MNRAS.519.1396S, 2023ApJ...954...70A},
has increasingly brought attention to hadronic models of blazars, particularly lepto-hadronic models \citep{2018ApJ...863L..10A, 2018ApJ...864...84K, 2018ApJ...865..124M, 2018MNRAS.480..192P, 2018ApJ...866..109S, 2019MNRAS.484.2067R, 2019MNRAS.483L..12C, 2019A&A...622A.144S, 2019NatAs...3...88G, GBS22, 2023MNRAS.519.1396S}.

The modeling of the broadband emission from blazars has always been of high
interest, as it provides crucial insights into the physics of jets. This includes
understanding their structure, the mechanisms of particle acceleration
within them, and the nature of their emissions, all crucial for advancing
our knowledge of relativistic jets. In recent years, progress in astrophysical
detectors and continuous multi-band monitoring of the sky have resulted in an
accumulation of extensive multiwavelength data from blazars.
This wealth of data not only enables the modeling of single
snapshot SED but also allows for systematic studies of blazar emission across
different periods \citep[see e.g.][]{2021MNRAS.504.5074S, 2022MNRAS.513.4645S,
2022MNRAS.517.2757S}. 

To interpret this wealth of data accurately and capture the many competitive mechanisms at work in blazar jets, numerical modeling is required. Consequently, various codes have been developed to model the lepto-hadronic processes
in blazar jets, such as naima \citep{naima}, JetSeT \citep{TGP09,TMT11,2020ascl.soft09001T},
agnpy \citep{2022A&A...660A..18N}, AM3 \citep{GPW17}, ATHE$\nu$A \citep{MC95}, Böttcher13
\citep{BRS13}, LeHa-Paris \citep{CZB15}, LeHaMoC \cite{SPV23}, and \textit{SOPRANO} \citep[Simulator
of Processes in Relativistic AstroNomical Objects, ][]{GBS22}. However, all these codes
encounter significant challenges in self-consistently fitting
blazar SEDs. These codes either rely on an ad-hoc
distribution of emitting electrons or face prohibitively long computation times that
does not permit thorough exploration of the parameter space
of the models. In this era of data-rich astrophysics, an accurate and comprehensive modeling
of blazar SEDs, which includes the detailed treatment of electron interaction and formation
of the electron spectrum, is essential for a deeper understanding of these complex astronomical
objects.

To address the significant computational challenges posed by traditional radiative models, we introduced an innovative approach for the self-consistent
modeling of blazar SEDs \citep{BSD23}. We employed a Convolutional Neural
Network (CNN) trained on a leptonic SSC model which accounts for processes such as electron cooling and pair creation-annihilation. Once trained, the network effectively reproduces the radiative signature
of the SSC model in a significantly reduced time and enables efficient modeling of the
broadband spectrum of BL Lacs thereby allowing to explore the parameter
space through fitting procedures. Shortly after, a similar study
using a different network approach was presented by \citet{2023arXiv231106181T}.

\citet{BSD23} is the first publication of the project titled "Modeling blazar broadband emission with convolutional neural networks." The primary goal of the project is to transform complex and computationally demanding lepto-hadronic radiative models into
efficient neural network frameworks. This transformation is aimed at enabling statistical
analysis and exploration of the parameter space governing blazar emissions. The simplicity of use and quick computation time also enable us to share the stand-in model directly with the blazar community through  Markarian
Multiwavelength Data Center (MMDC)\footnote{See \url{https://mmdc.am} under theoretical modeling}. The current paper
seeks to build upon the work we initiated in \citet{BSD23} by
developing a CNN dedicated
to an EIC model. Any EIC model presents additional challenges
due to its extra parameters, which are essential for characterizing the external 
photon fields. Our approach involves training the CNN on a dataset of spectra that are
numerically derived from an EIC model of blazars, using \textit{SOPRANO} \citep{GBS22}. By considering
a broad range of parameters for both the emitting electrons and the external
photon fields, this new CNN can be effectively
applied to study the emission from any FSRQ.

The paper is organized as follows. Section \ref{sec:model}
describes the EIC model, including the considered sources of the external photons.
The parameter ranges, simulation methodology and the implementation of the CNN are
detailed in Section \ref{sec:numerical_model}. The application of the developed
CNN to model the SEDs of 3C 454.3 and CTA 102 is presented in Section
\ref{sec:modeling}. Finally, the conclusions are given
in Section \ref{sec:conc}.

\section{The model: external inverse Compton}\label{sec:model}
\subsection{The kinetic model}

In this paper, we further develop the analysis technique
presented in \citet{BSD23} and consider the modeling of broadband SEDs of FSRQs. These SEDs are typically characterised by a pronounced 
Compton dominance, suggesting that, alongside synchrotron photons,
an external photon field significantly contributes to the formation of the HE and
VHE components. The source and nature of these external photons vary depending
on the emission region distance from the central black hole. Photons originating
directly from the black hole accretion disk,
those reflected from the BLR, or those emitted from the dusty torus, can all act
as potential targets for external Compton scattering by relativistic electrons
accelerated in the jet. This external inverse Compton mechanism has been effectively
used to model the observed multiwavelength data, as well as to explain various
features observed in narrow spectral bands, such as the blue bump in the optical
band, which is attributed to the accretion disc, etc. \citep[see e.g.][]{GTG09}.

In this model, the emission is assumed to originate from a
spherical blob with a radius of $R$, threaded by uniform
magnetic field. Further assuming that the jet half opening angle is $\sim 1/\Gamma$, $R$ is also the laboratory distance from the central black-hole to the blob. This blob moves with a Lorentz factor $\Gamma$ . As the
emission region is assumed to be oriented at a small angle relative to the
observer's frame, the emission is amplified by the relativistic Doppler
factor, denoted as $\delta$, which we assume to be such that $\delta = \Gamma$. The
electrons are injected into the emitting region as a power-law with an
exponential cutoff above the minimum Lorentz factor $\gamma_{\rm min}$, which is
the same distribution function we used for the SSC model \citep{BSD23}. This injection function is given by:
\begin{equation}
Q_e  = \left \{ \begin{aligned}
& Q_{e,0} \gamma^{-p} \exp \left ( -\frac{\gamma}{\gamma_{\max}} \right ) &  &    \gamma \geq \gamma_{\rm min}, \\
&0 & & {\rm otherwise.}
\end{aligned} \right.
\end{equation}
Here, $p$ represents the electron index, and $\gamma_{\rm max}$ is the electron
Lorentz factor beyond which electrons are not effectively accelerated. The
normalization $Q_{e,0}$ is used to scale the luminosity of the electrons:
\begin{equation}
L_e = \pi R^2 \delta^2 m_e c^3 \int_1^{\infty} \gamma Q_{e} d\gamma,
\end{equation}
where \( m_e \) denotes the electron mass.

The injected electrons interact with both internal
(synchrotron) and external photons, generating the HE component. To define
the spectrum of the external photons, we adopt the approach outlined by \citet{GT09}.
For clarity, the relevant definitions are provided below, but we refer to
\citet{GT09} for details. We consider three distinct external radiation fields, produced by (i) the accretion disk, (ii) the BLR, and (iii) the dusty torus. The contribution of each field scales with the distance of the emitting region from the central source. The accretion disk is modeled following the classical \citet{SS73} thin disk framework, where the disk temperature scales as:

\begin{align}
    T^4(R_d) = \frac{3 R_s L_d}{16\pi \eta \sigma_{SB} R_d^3} \left[ 1 - \left( \frac{3 R_s}{R_d}\right )^{\frac{1}{2}} \right ] 
    \label{eq:Tdisk}
\end{align}

In this equation, $R_s$ denotes the Schwarzschild radius, $\eta$ the
accretion efficiency, $\sigma_{SB}$ the Stefan-Boltzmann constant, and
$L_d = \eta \dot M$ represents the disk luminosity, with $\dot M$ being
the black hole mass accretion rate and $R_d$ is the radial distance from
the central black hole. The model includes several free parameters
describing the external radiation field from the disk, namely:
\begin{itemize}
    \item The disk luminosity $L_d$,
    \item The black hole mass $M$, inferred through the value of $R_s$,
    \item The radiative efficiency $\eta$, set to 0.1 following \citet{GT09},
    \item The inner and outer disk radii, set to $R_{\rm in} = 3 R_s$ and $R_{\rm out} = 500 R_s$. The outer radius has minimal impact as it does not significantly alter the external disk energy density, while the inner radius influences the maximum temperature.
\end{itemize}
Thus, incorporating the accretion disk's contribution into the radiative model
necessitates only two additional parameters, namely the disk luminosity $L_d$
and the black-hole mass $M$.

The BLR is modeled as a spherical shell located at a distance
$R_{\rm BLR} = 10^{17} \, \text{cm} \left( L_d/10^{45} \, \text{erg s}^{-1} \right)^{1/2}$
\citep{GT09}. This region reflects a fraction \( f_{\rm BLR}=0.1 \) of the disk luminosity.
The spectrum of the BLR is assumed to be a black body, peaking at a frequency of
\( \nu_{\rm BLR} = 2.47 \times 10^{15} \) Hz \citep{GT09}. Similarly, the dusty torus
is assumed to be as a spherical shell located at a distance
$R_{\rm DT} = 2.5 \times 10^{18} \, \text{cm} \left( L_d/10^{45} \,
\text{erg s}^{-1} \right)^{1/2}$ which emits a black-body
spectrum peaking at \( \nu_{\rm IR} = 3 \times 10^{13} \) Hz and reprocesses a fraction
\( f_{\rm DT}= 0.5 \) of the disk
luminosity. To leave additional freedom to the model, we
elected to allow the temperatures of the BLR and of the dusty torus to vary within limited ranges around
the proposed values. It is important to note that these temperatures are
measured in the frame of the central supermassive black hole, and are Lorentz transformed
(meaning boosted) to the frame comoving with the blob when used in the computation.

The temporal evolution of the electron distribution is determined by considering
the interaction of electrons with the magnetic field and all
photon fields, namely of external (from the accretion disk, the dusty torus
and the BLR) and internal (all leptonic processes taking place in the jet) origins.
We denote the distribution functions of electrons and photons as $N_e$ and $N_\gamma$,
respectively, and the distribution function of the external photon fields as
$N_{\gamma, \rm ext}$, such that $N_{\gamma, \rm ext}$ is the sum of the boosted spectra
from the three external sources considered here. The photon energy and the Lorentz
factor of the electrons are represented by $x$ and $\gamma$, respectively. The kinetic
equation that describes the time evolution of the electrons is approximated by the
Fokker-Planck diffusion equation, while the equation for the photons is an
integro-differential equation. These are expressed as follows:
\begin{align}
\left \{  \begin{aligned}
    \frac{\partial N_e}{\partial t} (\gamma ) &= Q_e + \frac{N_e}{t_{\rm esc}} + \frac{\partial}{\partial \gamma } \left \{ N_e \times [ C_{\rm IC}(N_\gamma + N_{\gamma, \rm ext} ) + C_{\rm sync}  ] \right \} + Q_{\gamma \gamma \rightarrow e^+e^- } (N_\gamma + N_{\gamma, \rm ext})   \\
    \frac{\partial N_\gamma}{\partial t} (x) &= \frac{N_\gamma}{t_{\rm esc}} + Q_{\rm  sync}(N_e) + R_{\rm IC}(N_e, N_{\gamma, \rm ext}) - S_{\gamma \gamma \rightarrow e^+e^-} (N_\gamma + N_{\gamma, \rm ext})
\end{aligned} \right. \label{eq:kinetic_equation}
\end{align}
Here, $t_{\rm esc} \equiv t_{\rm dyn} = R/c$ represents the escape time, such that the first term on the
right-hand side of each equation represents the escape of particles from the
radiation zone. $C_{\rm IC} $ and $C_{\rm sync}$ denote the inverse Compton and
synchrotron cooling, respectively. Terms labeled $Q$ and $S$ are source and sink
terms, and $R_{\rm IC}$ is the redistribution kernel of Compton scattering.
Distribution functions in parenthesis represent the functional
dependence. Further details on the kinetic equations and their numerical
solutions, along with the expressions for all the rates that appear in this
set of equations, can be found in \citet{GBS22}.

The set of coupled kinetic equations given by Equation \ref{eq:kinetic_equation}
is solved using \textit{SOPRANO} \citep{GBS22} which is based on the finite volume method
for the discretization in energy and uses a semi-implicit numerical scheme, in the
sense that leptonic interaction are treated in a fully implicit framework, while
the target of hadronic interactions are treated explicitly. Since there
is no hadronic component in the model used in this paper, the time integration is
fully implicit. The time dependent nature of \textit{SOPRANO} permits
in depth modeling of the observed multimessenger SEDs, for application of
\textit{SOPRANO} see e.g. \citet{GBS22, 2023ApJS..266...37A} and \citet{2023MNRAS.519.1396S}.
For the need of the current model, we assume that the system can reach equilibrium,
and its evolution becomes independent on time. In order to find the
equilibrium solution to the kinetic equations, we evolve the solution until
time $t = 4 t_{\rm dyn}$ after which we find that the variations of the spectrum
are too small to be of any interest. As a results, the outputs of \textit{SOPRANO}
which are used to create the model database to be used as the training set.

\begin{table*}
\centering
\begin{tabular}{|c|c|c|c|c|c|}
    \hline
     Parameters & Units  & Symbol  & Minimum & Maximum & Type of distribution \\ \hhline{|=|=|=|=|=|=|}
     Doppler boost & -  & $\delta$ & 3 & 50 & Linear \\ 
     Blob radius        & cm &     R    &  $10^{15}$   &  $10^{18}$  &  Logarithmic \\
     \makecell{Minimum electron Lorentz factor} & - & $\gamma_{\rm min}$ & $10^{1.5}$ &  $10^5$ &  Logarithmic \\
     \makecell{Maximum electron Lorentz factor} & - & $\gamma_{\rm max}$ & $10^2$ &  $10^6$ & Logarithmic \\
     Injection index & - & $p$ & $1.8$ & 5 & Linear \\
     Electron luminosity & erg.s$^{-1}$ & $L_e$ & $10^{42}$ & $10^{48}$ &  Logarithmic \\
     Magnetic field  &  G  & $B$  & $10^{-3}$  & $10^{2.5}$ & Logarithmic  \\ 
     Black hole mass & $M_\sun$ & $ M_{\rm BH}$ & $10^7$ & $10^{10}$ & Logarithmic  \\
     Disk luminosity & erg.s$^{-1}$ & $L_d$ & $10^{43.5}$ & $10^{47.5} $ & Logarithmic  \\
     BLR frequency   &  Hz &  $\nu_{\rm BLR}$ & $10^{14.5}$ & $10^{16}$ & Logarithmic  \\ 
     DT frequency   &  Hz &  $\nu_{\rm DT}$ & $10^{12.5}$ & $10^{14}$ & Logarithmic  \\ \hline
\end{tabular}
    \caption{Properties of the dataset used in training the CNN. This includes the unit (where applicable) and symbol for each parameter, along with its range and the distribution across discrete parameter values. The dataset contains a total of $10^6$ spectra.} 
    \label{tab:table_parameters}
\end{table*}

\section{Numerical model: computation, validation and CNN} \label{sec:numerical_model}

To effectively train the CNN, it is essential to have a substantial number of SEDs for
a wide range of parameters. Unlike the SSC model discussed in \citet{BSD23}, the EC model
exhibits significant spectral variation,
so  a larger dataset of training SEDs is needed for the CNN to achieve adequate accuracy.
To address this, \(10^6\) SEDs were generated using \textit{SOPRANO}, to be compared to the $2\times 10^5$ spectra for the SSC model of \citep{BSD23} and the $10^4$ spectra of \citet{2023arXiv231106181T}.

\subsection{Parameter ranges and sampling}

In the EC model considered in this study, there are eleven free parameters,
namely the comoving blob radius $R$, the Doppler factor of the emission region
$\delta$, the comoving magnetic field strength $B$ within the emission zone,
the electron luminosity $L_e$, the minimum Lorentz factor $\gamma_{\text{min}}$,
the cutoff Lorentz factor $\gamma_{\text{max}}$, the power-law index $p$, the
accretion disk luminosity $L_d$, the mass of the central supermassive black-hole
$M_{\rm BH}$, and the temperature/frequency of the broad line and dusty torus.
These parameters, along with their respective units and ranges, are given in
Table \ref{tab:table_parameters}. The parameters characterizing the emission
region and the electron distribution functions are consistent with those used
for the SSC modeling in \citet{BSD23}.
The Doppler boost factor linearly spans a range from 3 to 50 and  the power-law
index $p$ is linearly sampled between 1.8 and 5. It is important to note, however,
that steep values of $p > 3$ are generally not anticipated from
the theory of shock acceleration or magnetic reconnection, as discussed in e.g. \citet{KGG00, SS11, Uzd22}. Despite these
theoretical considerations, we have chosen to explore
a broader range of values of $p$ which allow us to
comprehensively investigate the parameter space without being constrained by its
boundaries. In the fits, setting the index to a given value is straightforward, and priors can be used to reduce its range.

The remaining  model parameters, \textit{i.e} the emission radius $R$,
the minimum and maximum Lorentz factors $\gamma_{\rm min}$ and $\gamma_{\rm max}$,
the electron luminosity $L_e$, the strength of the comoving magnetic field $B$, the
mass of the central black-hole $M_{\rm BH}$, the disk luminosity $L_d$, the temperature
of the broad line region and of the dusty torus are all sampled logarithmically within
their respective ranges. Specifically, these ranges are given by $15 < \log(R) < 18$,
$ 1.5 < \log(\gamma_{\rm min}) < 5 $, $ 2 < \log(\gamma_{\rm max}) < 8 $,
$ 42 < \log(L_e) < 48 $, $ -3 < \log(B) < 2 $, $ 7 < \log(M_{\rm BH}) < 10 $,
$ 43.5 < \log(L_d) < 47.5 $, $ 14.5 < \log(\nu_{\rm BLR}) < 16 $ and
$ 12.5 < \log(\nu_{\rm DT}) < 14 $.

This large range of the parameters ensures that
the CNN we developed will be applicable for the modeling of
any FSRQ SED.
In addition, the low achievable values for the disk luminosity guaranties that the trained model
can also be used to model BL Lacs, leaving external fields unconstrained. In other words, our approach allows
for immediate model comparison between the SSC and EC model.
We employ \textit{ronswanson} \citep{Bur23} for
generating the set of parameters to be simulated with \textit{SOPRANO}. The parameters are distributed following the  Latin hypercube
sampling technique, which provides an
efficient way to cover a multidimensional parameter space with uniform parameter
distributions \citep[see][]{MBC00,Via16}. This approach ensures a thorough, uniform and
representative sampling of the parameter space.

Finally, in the generation of the SEDs with \textit{SOPRANO}, the accuracy threshold for the Newton-Raphson
method was set to \(10^{-10}\), which is five orders of magnitude larger than the threshold
used for the SSC model in \citet{BSD23}. Provided the large number of spectra that needed to be computed, this adjustment was made to achieve a significant reduction of the compute time per spectra.
%}

\subsection{Computation of the spectra on Amazon Elastic Computing Cloud}

\begin{figure}
    \centering
    \includegraphics[width = 0.95 \textwidth]{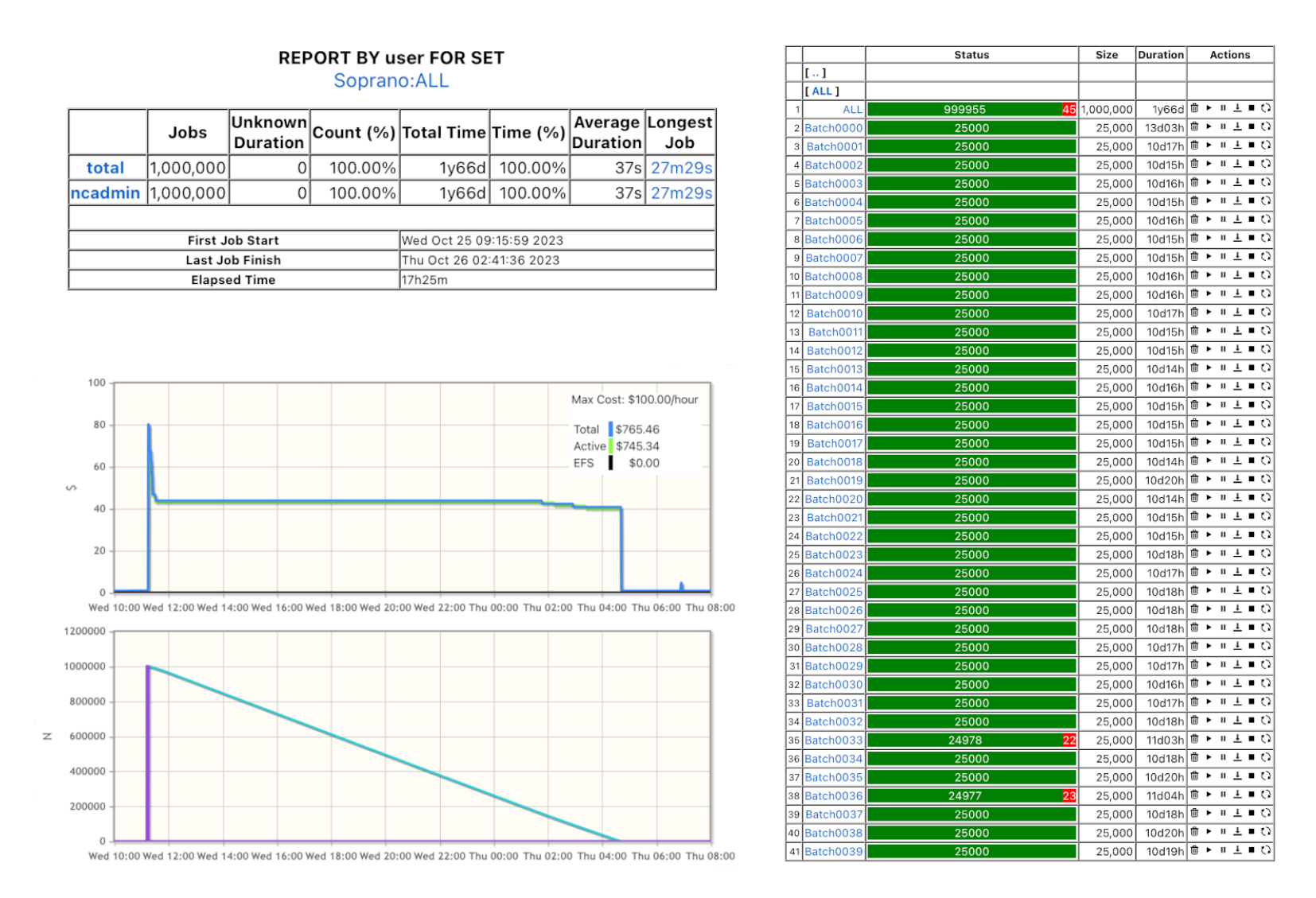} \\
    \caption{FNC data handling. {\it Top left:} A summary view of the workload, generated by FNC. FNC handled one million jobs corresponding to the computation of each individual spectra for a total compute time of 1 year and 66 days, performed in only 17 hours and a half. {\it Bottom left:} The top graph shows the total cost incurred for this computation as a function of time.  The bottom graph shows the jobs queued waiting to be executed. {\it Right:} Report showing the workload, for convenience broken down into 40 sets of 25,000 jobs each.  Notice that 45 jobs have been labeled as failed after the first computation, due to the interruption of spot instances. FNC allows effortless rescheduling of those 45 failed jobs and within a few minutes all jobs were done.}
    \label{fig:fnccost}
\end{figure}

The core numerical workload consists of
the generation of $10^6$ spectra, where each spectrum
is computed independently. This results in 1 million \textit{independent} jobs, 
with each job executing a Python script.
Every job requires precisely 8 cores and runs on average for 37s, with a maximum of
over 27m (see top left panel of Figure \ref{fig:fnccost}).This duration differs
from the runtime of \textit{SOPRANO} which does not account for the loading of the
cross-section and for the initial setup. The peak RAM usage is slightly above 10GB.
Executing this workload on a single 8-core machine would take approximately 14 months, so it is imperative to run the workload on multiple machines. We note that we estimated the total compute time beforehand by computing two smaller databases with 500 and 5000 spectra respectively. This step also allowed us to estimate the relevance of the parameter ranges and foresee any computational bottleneck in this large parameter space.

Although we have access to a small server with shared resources among researchers, managing a large number of independent jobs efficiently presents many challenges. It would require us to perform the computation of the spectra in large bunches via the addition of a management layer, which also be able to keep track of computation status and failure. All of this at the price of monopolizing the compute resources for extended periods of time: even with this approach, we would still face several months of computation.

We have consequently transferred the workload to Amazon Elastic Computing Cloud (AWS EC2). For the workload manager, we used FNC, developed by Altair Labs, for more details see Casotto (2024, in prep.)\footnote{Additional information can also be found on the website of Altair: \url{https://altair.com/newsroom/articles/phoenix-rising-next-generation-job-scheduling-with-fenice}}. FNC was chosen for the following reasons:
\begin{itemize}
\item its capability to handle several million jobs, contrasting with common schedulers. FNC maximizes the resource usage while minimizing the cost of the numerical computation;
\item it keeps the state of every independent jobs even after their completion; 
\item it includes a ``Rapid Scaling'' patented module that can grow and shrink the farm depending on the workload;
\item A browser-based interface that facilitates easy monitoring of the workload and real-time costs associated with the rental of Amazon servers.
\end{itemize}

As head node, we deployed a 4-core, 32GB machine of type t3.xlarge. For computing, AWS EC2 offers two types of instances:

\begin{enumerate}
\item On-demand instances, which remain available for as long as needed 
but which are substantially expensive;

\item Spot instances, which are offered at a discount relative to
on-demand instances, with the stipulation that the instance could
be revoked at any time by EC2 if it decides to give such resources
to a customer willing to pay the full price.
\end{enumerate}

We selected two
instance families, namely  c7a and  r7i, based mainly on the spot cost at
the time of the computation, which was about \$0.06/hour
for an 8-core instance (like c7a.2xlarge). Given that the runtime of each job
is short (on average 37s), the computation was performed on spot instances, as the cost penalty for resuming an interrupted job is small.
  
Upon submission of the workload,
the Rapid Scaling module grabs available spot instances from AWS EC2
from the preferred families. We set a limit of 600 concurrent jobs, for a
total max size of the farm of $8 \times 600 = 4800$ cores. Towards the end of the computation, the Rapid Scaling module
automatically returns an instance to the farm if this instance
has been idle for more than 1 minute, thereby enabling saving money
by releasing non-used compute instances. 

In bottom left panel of Figure \ref{fig:fnccost}, we show the time evolution
of the total price rate (in \$/hour) incurred for this computation, resulting in a total expense of about \$767. 
The peak cost of \$80/hour at the beginning of the computation was caused by the fact that Rapid Scaling initially estimated that
1 million queued jobs are in need of more hardware, so it tried
to get as many instances as it can. Shortly after, that is when 600 concurrent jobs are
running, the workload is limited by the self-imposed policy on the number of
concurrent jobs,  idle instances are rapidly terminated to reduce the incurred cost. 

The entire workload was completed in just over 17 hours.
The FNC workload manager detected that 45 out of $10^6$ jobs
failed (see right panel of Figure \ref{fig:fnccost}), because of the sudden termination of a couple of spot
instances. This is an inherent risk of using spot instances, which in our case
is fully acceptable since the computation of individual spectra are short. Given
that FNC maintains the state of the jobs even after completion, it was
a matter of a couple of clicks in the browser based interface to
re-execute the 45 failed jobs.

\begin{figure}
    \centering
    \begin{tabular}{cc}
    \includegraphics[width = 0.45 \textwidth]{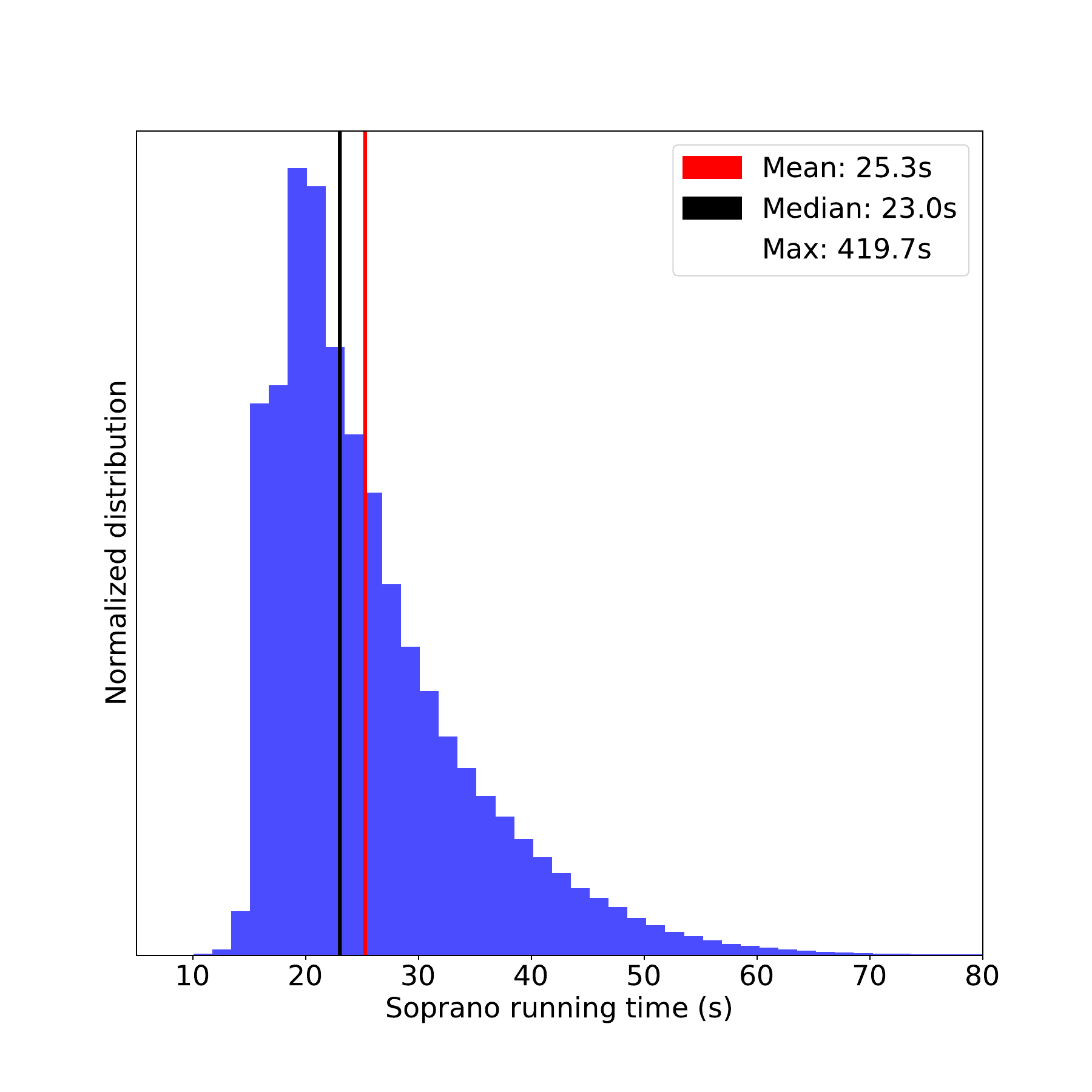}  & 
    \includegraphics[width = 0.45 \textwidth]{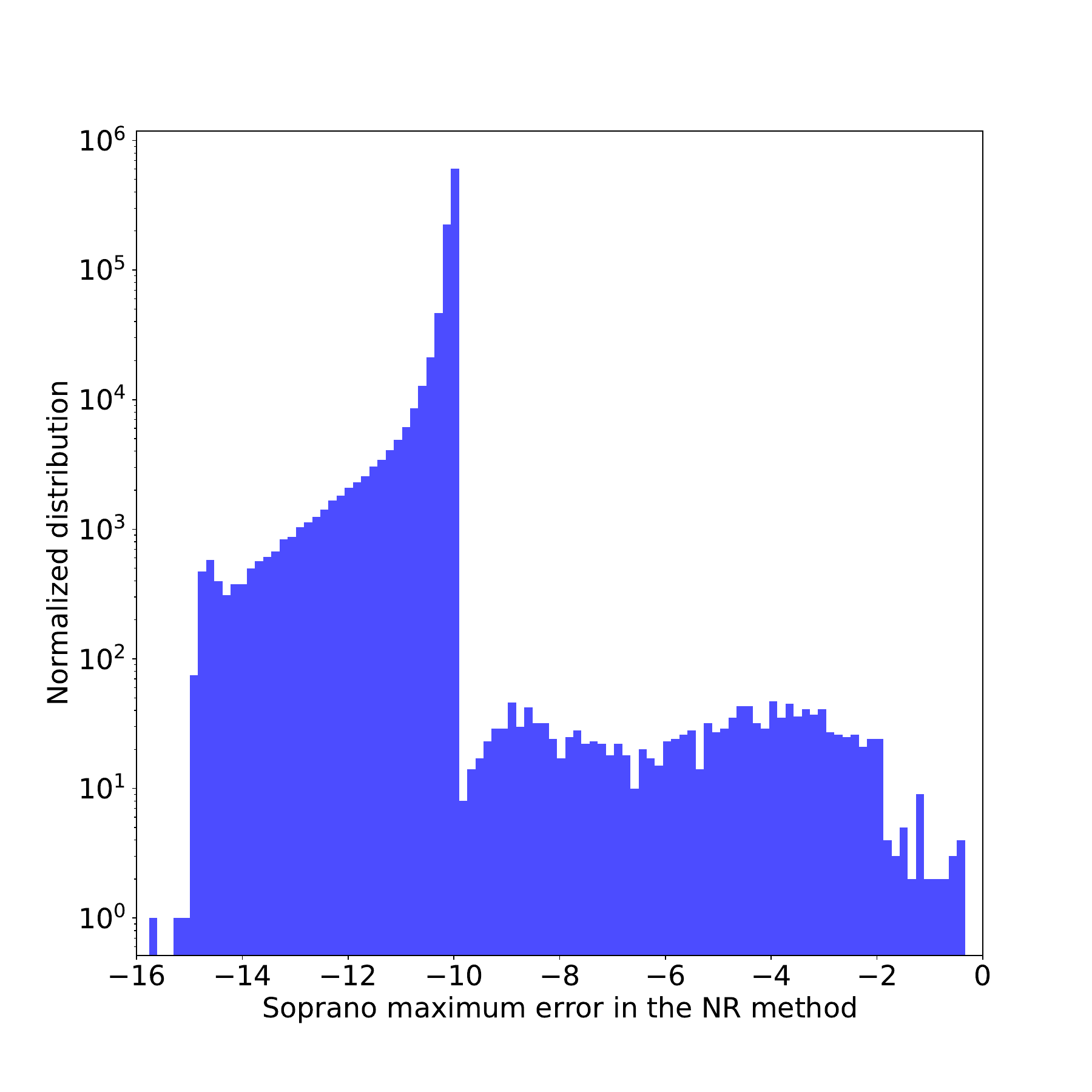}
    \end{tabular}
    \caption{Left: distribution of the run time for each simulations, each using 8 cores. Right: Maximum error over the course of each simulation. Most simulations have errors below the targeted accuracy.}
    \label{fig:meta_data}
\end{figure}

\subsection{Convolutional Neural Network}

\label{sec:CNN}

Once the computation is finished, the meta data are analysed.
The distribution of computation time per spectra and of the maximum error of the
Newton-Raphson scheme are displayed in Figure \ref{fig:meta_data}. First,
lowering the accuracy of the Newton-Raphson error from $10^{-15}$ to $10^{-10}$ resulted
in a significant reduction in computation time for the spectra, averaging
25.3 seconds compared to 43.7 seconds for the SSC model, despite the increased
complexity of the EC model. Additionally, this approach led to a lower
fraction of failed spectra $\sim 0.15$ \%, defined as those not meeting the
required accuracy at least once during computation. Based on these results, we conclude that
the majority of the spectra computed with \textit{SOPRANO} are sufficiently
accurate to be utilized as a training sample for our CNN.
 
We employ the same CNN and optimization methodology as those used in \citet{BSD23}.
Specifically, we incorporate numerical derivatives into the training output,
to smooth out the spectra computed with the CNN. As highlighted in \citet{BSD23},
this technique also improves the convergence rate. In order to obtain the most
accurate CNN, the model was trained 16 times, with varying batch sizes,
training fractions, and random distribution between training
and validation sets. The optimal CNN was then selected based on
its performances evaluated from its R2 score and mean squared error. The contribution
of the external field to the observer is computed semi-analytically for each parameter set. Consequently, the total
spectrum used in the fit consists of the sum of the spectrum computed by the CNN and the contribution from the external field.

\begin{figure*}
    \centering
    \includegraphics[width=0.95\textwidth]{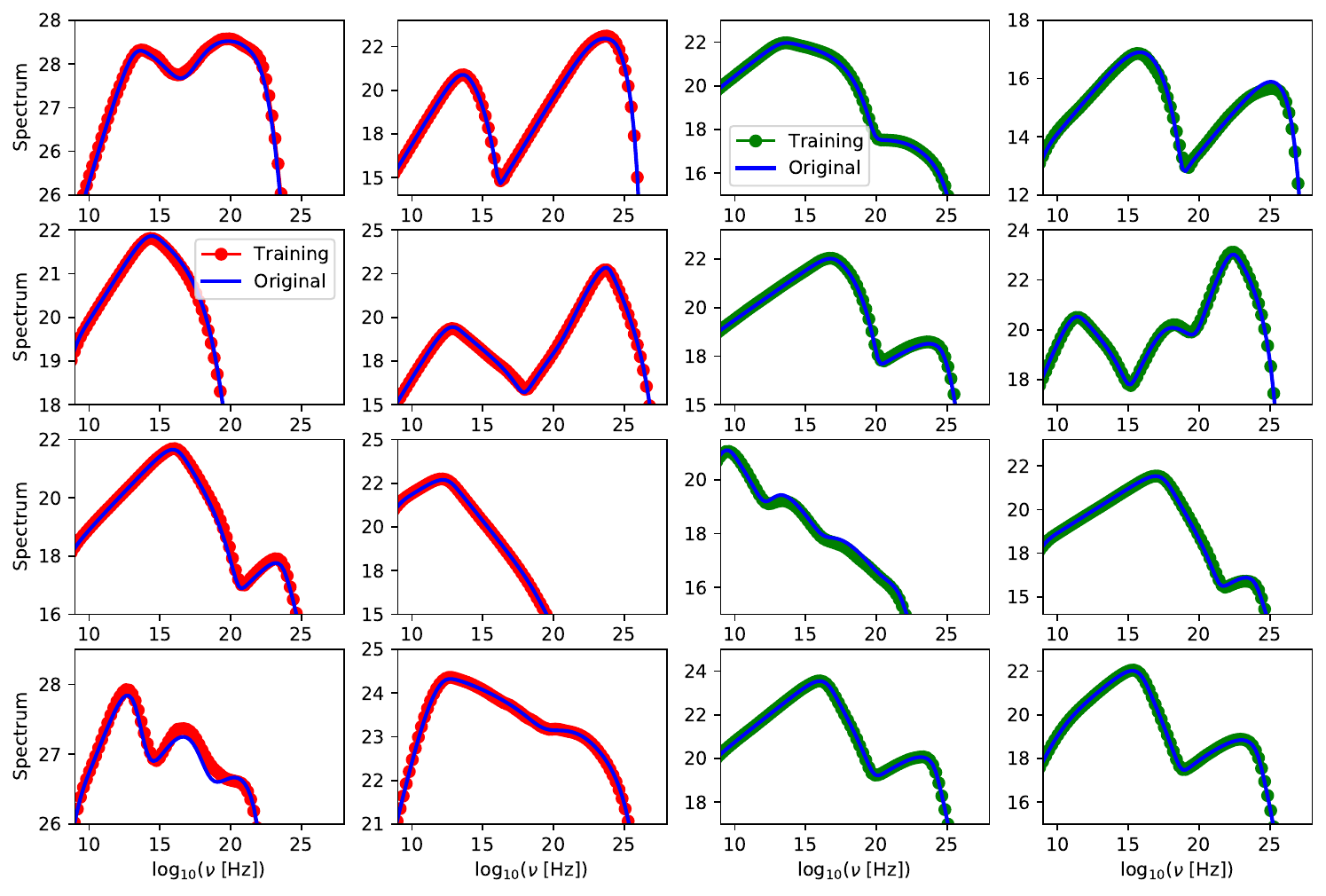}
    \caption{Comparison between the SED ($\nu$ versus $\nu F_\nu$) as calculated by the CNN — depicted with dots — against those generated by \textit{SOPRANO}, represented by a solid line, prior to conversion into the observer's frame and unit normalization. The left panel illustrates spectra derived from the training dataset, while the right panel displays spectra from the validation dataset. The diversity in spectra that the CNN can and does accurately reproduce is highlighted, along with the broad range of typical emitted powers predicted by the EC model.}
    \label{fig:CNN_spectra}
\end{figure*}

In Figure \ref{fig:CNN_spectra}, we present examples of spectra generated by
\textit{SOPRANO} alongside the corresponding results from the CNN for both the training
set (on the two leftmost columns) and the validation set (on the two rightmost
columns). The CNN demonstrates high performances with notable agreement across
all spectra between the original \textit{SOPRANO} outputs and the CNN results. This
performance is further evidenced by the following metrics: the model achieved an
averaged $R^2$ score of 0.87, a root mean-squared error (RMSE) of $6.01\times 10^{-5}$,
a mean-absolute error (MAE) of $1.67 \times 10^{-3}$, and a criterion
of $6.01 \times 10^{-5}$.

We also developed a CNN for the electron distribution function, aiming at
displaying the uncertainty on the comoving distribution function of electrons.
The main difference with the training of the photon spectrum is that the
training is performed on $\gamma^2 N_e$ rather than on $N_e$. We find
that this produces more accurate results, particularly at the transition
between electron cooling states, namely between the slow and fast cooling regimes.
While it might have been possible to further optimize the depth and the size of the network to produce
acceptable results with $N_e$ instead of $\gamma^2 N_e$, this would 
necessitate a large investment in work and in computation time.
This is not essential for the current purpose:
to depict the steady state electron distribution
function, for which a rather low accuracy is sufficient.

\section{Application: Modeling multiwavelength emission of 3C 454 and CTA 102}\label{sec:modeling}

The newly developed CNN presented in this study is used
to model the multiwavelength SEDs of two FSRQs, namely 3C 454.3 and CTA 102. This approach follows the
methodology of \citet{BSD23} in determining the parameters for the emitting electrons,
specifically their power-law index \(p\), luminosity \(L_{e}\),
\(\gamma_{\rm min}\), and \(\gamma_{\rm max}\), as well as the parameters
of the emitting region, such as \(\delta\), \(B\), and \(R\). 
We adopt a Gaussian likelihood to sample the posterior distributions using the MultiNest
algorithm \citep{FHB09}. For effective sampling and convergence, we utilize 1000 active points and set
the tolerance at 0.5. We note that the extragalactic background light (EBL) absorption
is taken into account via the model of \citet{2011MNRAS.410.2556D}.
\begin{table}
    \centering
\caption{Parameters describing the SEDs in Fig. \ref{sed}: The first column corresponds to the modeling of 3C 454.3, where all parameters of the emitting electrons, emission region, and $L_{\rm d}$ were considered free. In the modeling of CTA 102 (last column), $\gamma_{\rm min}=100$ is assumed. We set \( \nu_{\rm BLR} = 2.47 \times 10^{15} \) Hz and \( \nu_{\rm IR} = 3 \times 10^{13} \) Hz, and use \(1.5 \times 10^9 M_\odot\) and \(8.5 \times 10^8\: {\rm M_{BH}}\) for the black hole masses of 3C 454.3 and CTA 102, respectively.}
\label{tab:param}
    \begin{tabular}{c||c||c}
    \hline
    Parameters  &  3C 454.3 &  CTA 102  \\\hline 
    \( p \) & \( 1.87 \pm 0.18 \) & \( 1.97 \pm 0.27 \) \\
    \( \log_{10}(\gamma_{\rm max}) \) & \( 3.30 \pm 0.16 \) & \( 3.18 \pm 0.44 \) \\
    \( \log_{10}(\gamma_{\rm min}) \) & \( 1.90 \pm 0.10 \) & 2 \\
    \( \delta \) & \( 49.14 \pm 2.53 \) & \( 40.74 \pm 8.32 \) \\
    \( \log_{10}(B/[\rm G]) \) & \( -0.70 \pm 0.35 \) & \(  -0.06 \pm 0.85 \) \\
    \( \log_{10}(R/[\rm cm]) \) & \( 16.13 \pm 0.37 \) & \( 15.67 \pm 0.93 \) \\
    ~~~\( \log_{10}(L_{\rm e}/[\rm erg\:s^{-1}]) \)~~~ & ~~~ \( 46.75 \pm 0.44 \) ~~~ & ~~~ \( 45.95 \pm 0.64 \) ~~~  \\\
     ~~~\( \log_{10}(L_{\rm d}/[\rm erg\:s^{-1}]) \)~~~ & ~~~ \( 47.11 \pm 0.16 \) ~~~ & ~~~ \( 45.70 \pm 0.96 \)  ~~~ \\\hline
    \(\log_{10}(L_{\rm B}/[\rm erg\:s^{-1}]) \) & \( 43.82 \) & \( 44.02 \) \\
%    \hline
%      &  All parameters free & All parameters free & Variability time constraint
  \end{tabular}
  \label{params}
\end{table}

\subsection{3C 454.3} 

3C 454.3, a classical FSRQ located at a redshift of \(z = 0.859\), hosts a central black
hole with an estimated mass of \(1.5 \times 10^9 M_\odot\) \citep{2002ApJ...579..530W,
2006ApJ...637..669L}. This source is known for its frequent flaring across various spectral
bands, making it a frequent target for multiwavelength studies and resulting in a
substantial accumulation of data. The flares of 3C 454.3 are characterized not only by
a flux increase but also by complex changes both within and between different bands. For instance, during the high state in July 2008, the source's
emissions in the IR, optical, UV, and \gray\ bands,
were correlated while the evolution of the X-ray emission was independent \citep{2009ApJ...697L..81B}. Conversely,
during the extreme brightening in December 2009, a correlation
between the optical, X-ray, and \gray\ bands was observed
\citep{2011MNRAS.410..368B}. The source exhibited significant flares in the \gray\ band,
where the flux has exceeded approximately \(10^{-5}\) photons cm\(^{-2}\)
s\(^{-1}\) \citep{2011ApJ...733L..26A, 2021MNRAS.504.5074S}, which, at the distance
of 3C 454.3, corresponds to an isotropic luminosity of about \(10^{50}\) erg s\(^{-1}\)
in \grays. This luminosity ranks 3C 454.3 among the brightest blazars observed.

\begin{figure*}
    \centering
    \includegraphics[width=0.48\textwidth]{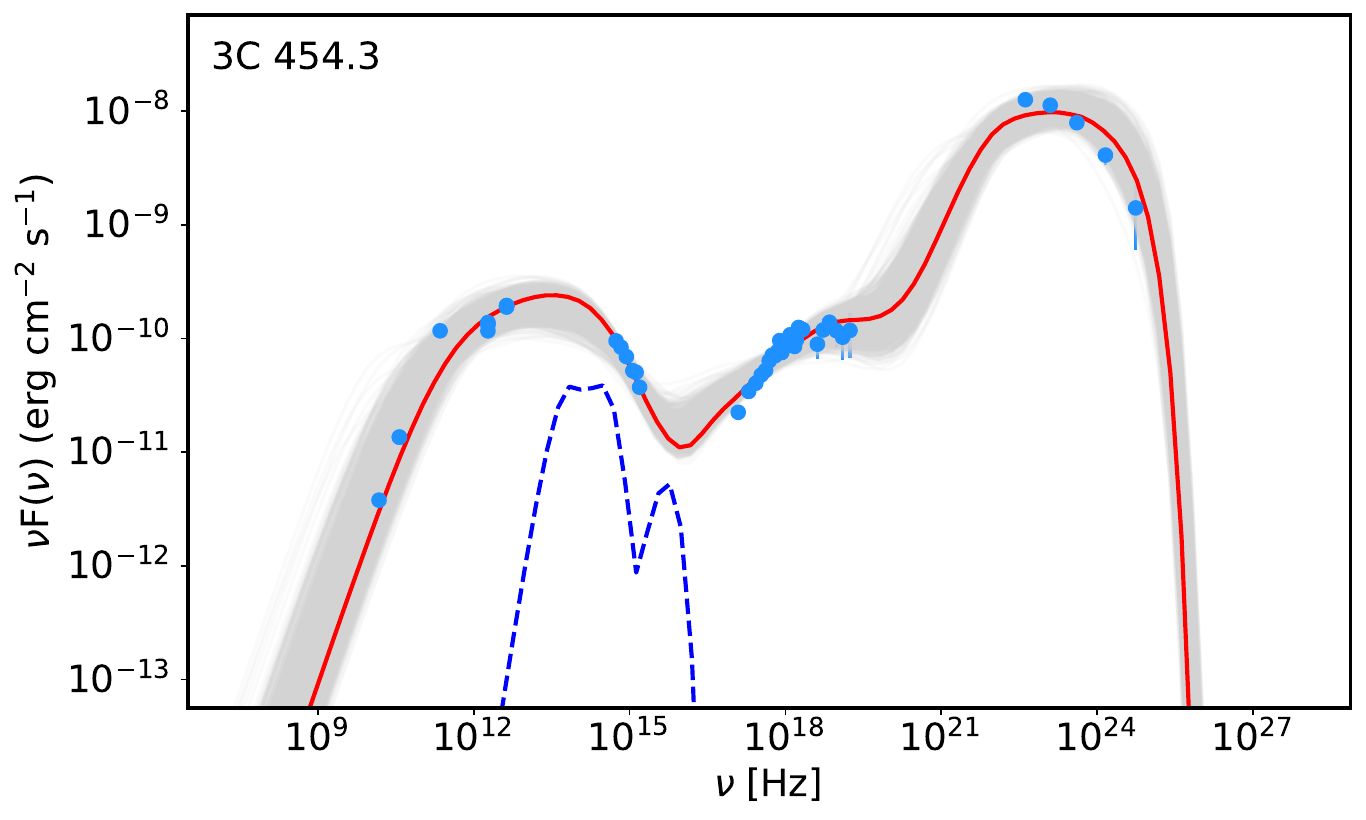}
     \includegraphics[width=0.48\textwidth]{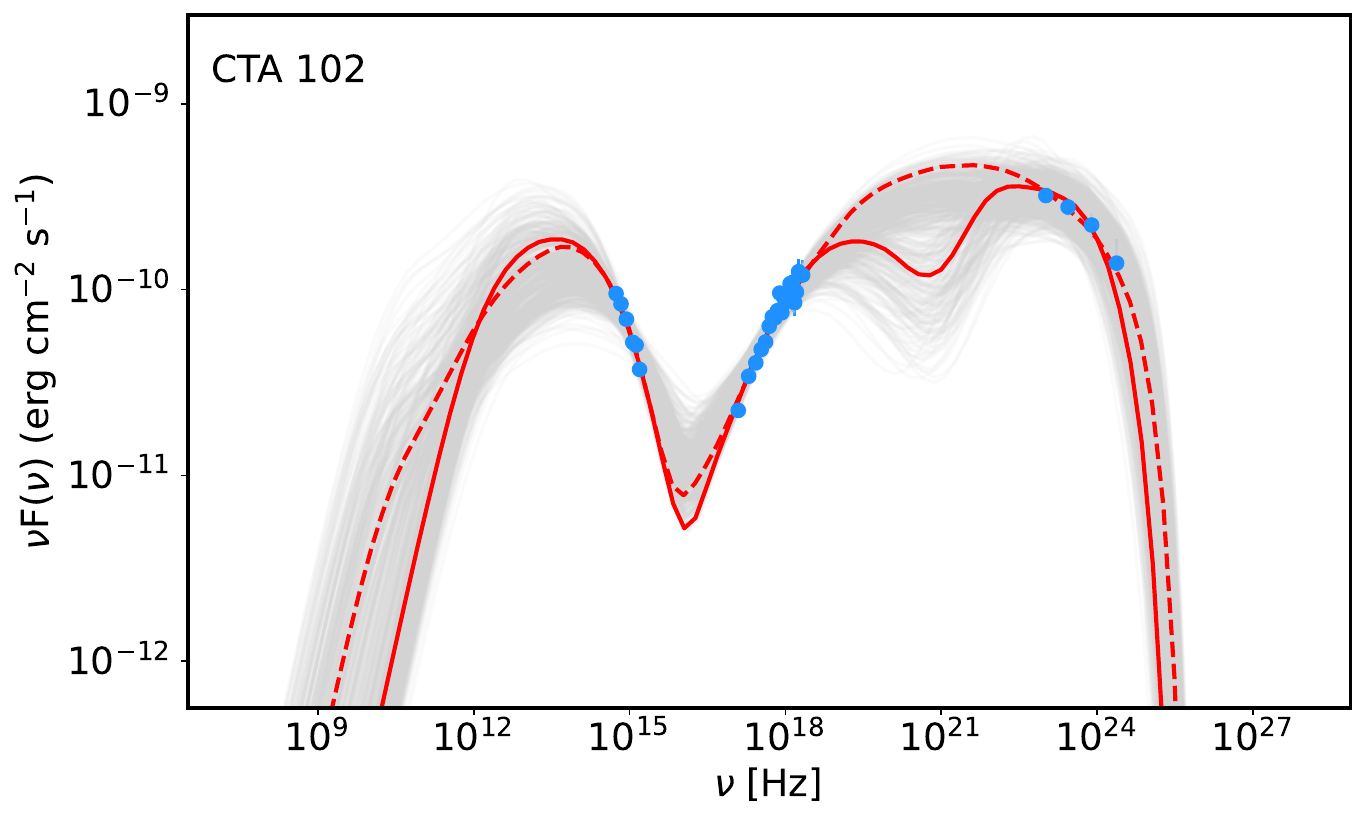}\\
     \includegraphics[width=0.48\textwidth]{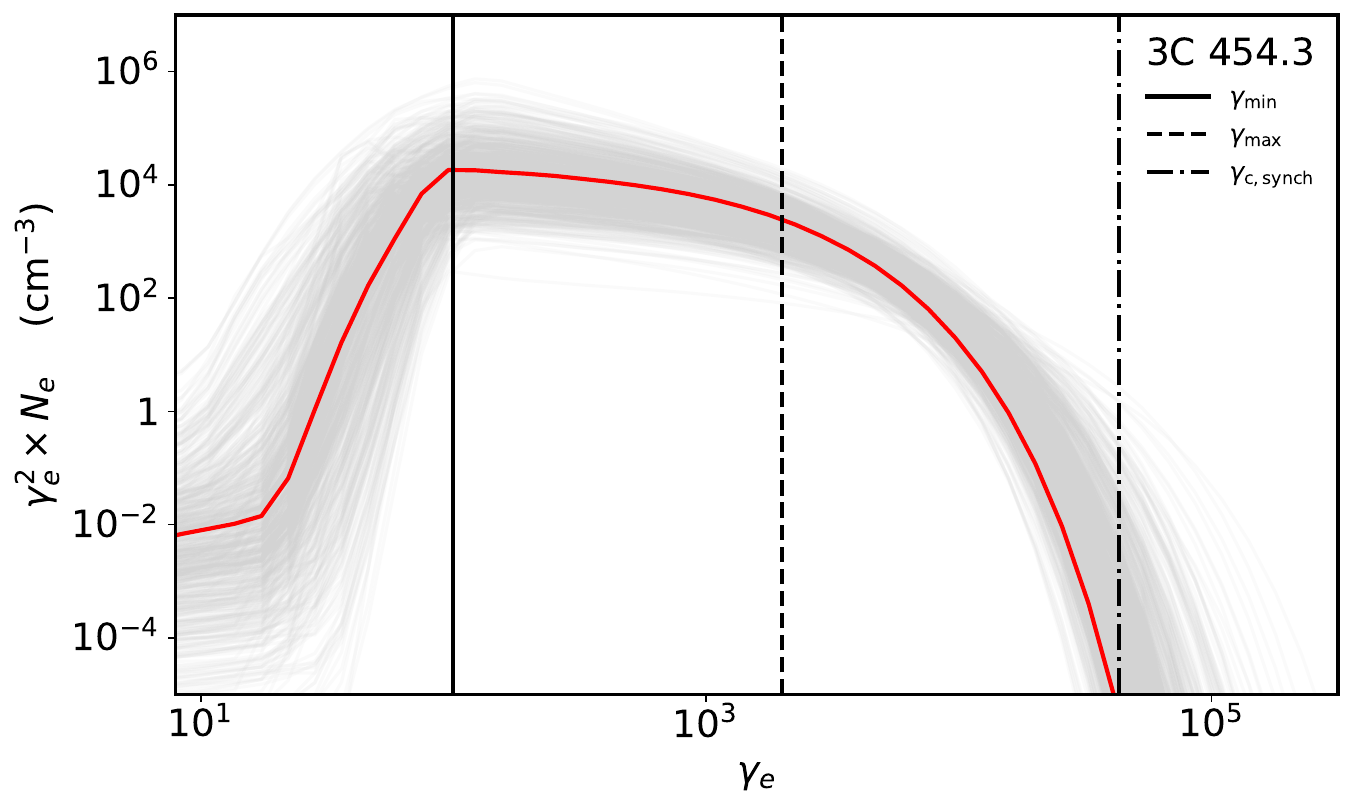}
     \includegraphics[width=0.48\textwidth]{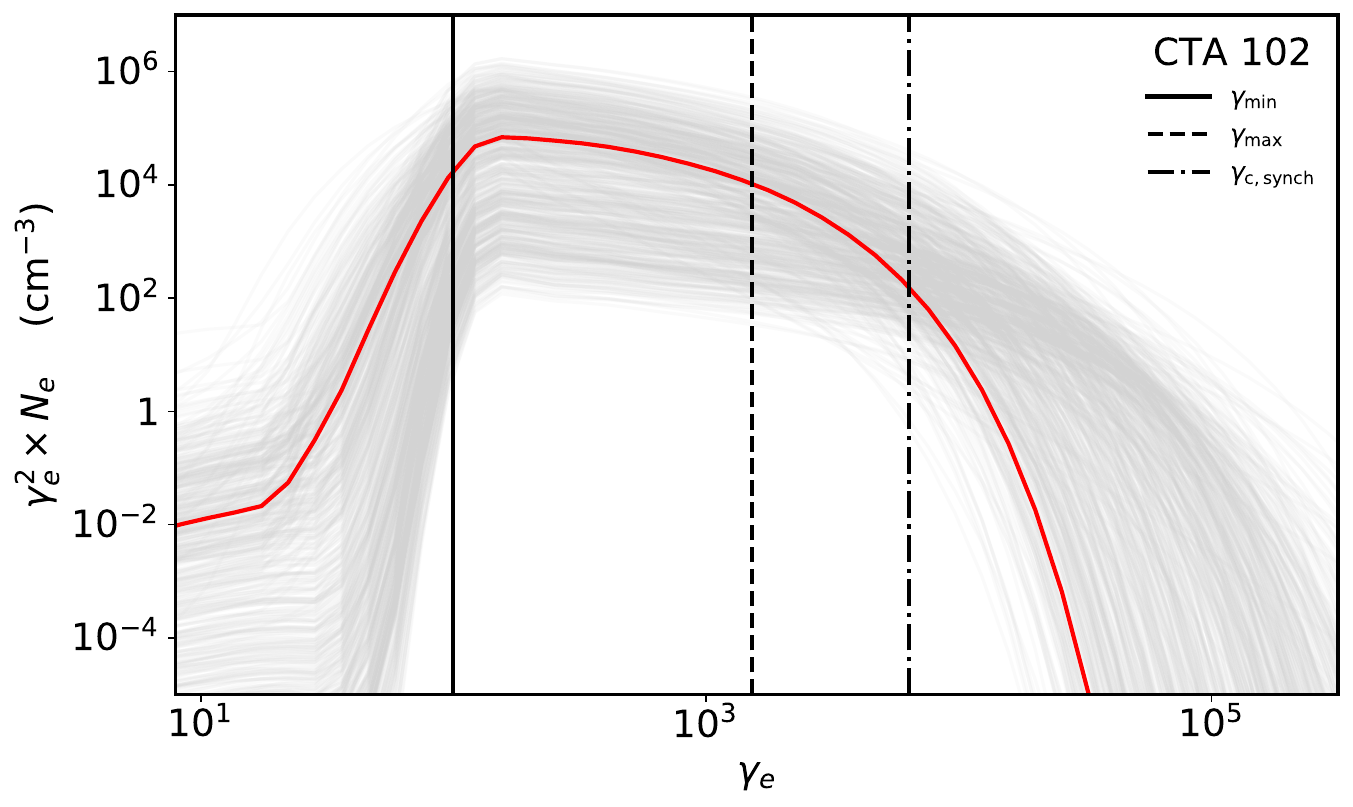}
    \caption{{\it Upper panels:} Multiwavelength SEDs of 3C 454.3 (left) and CTA 102 (right), presented
    in blue. On each panel, the model corresponding to the maximum
    likelihood is shown by the solid red line, while the uncertainty associated with the model is
    depicted in gray. In the left panel, the dashed blue line represents photons from the disk (low-energy peak) and the BLR. In the right panel, the red dashed line represents the scenario for which the radius is larger and the emission is due to SSC.
   For both blazars, the models account for EBL absorption, in other words, the data are not EBL corrected. {\it Lower panels:} Comoving electron distribution function at equilibrium for 3C454 (left) and CTA 102 (right) corresponding to the SED shown above. The red line corresponds to the best-fit parameters, while the grey lines represent the uncertainty. The vertical black lines refer to the characteristic electron Lorentz factor $\gamma_{\rm min}$, $\gamma_{\rm max}$ and $\gamma_{\rm c, ~synch}$ for the best model parameters.}
    \label{sed}
\end{figure*}

The broadband SED of 3C 454.3 for the period MJD 55519.59-55520.19, retrieved
from \citet{2021MNRAS.504.5074S}, is depicted on the left of
Figure \ref{sed}. This selected period corresponds to an active state of 3C 454.3, characterized by increased emission across all considered bands. In the SED modeling, it is
possible to vary all model parameters, but for our analysis, we have fixed the
black hole mass at $M_{\rm BH}=1.5 \times 10^9 M_\odot$, the BLR and dusty torus temperatures at
\( \nu_{\rm BLR} = 2.47 \times 10^{15} \) Hz and \( \nu_{\rm IR} = 3 \times 10^{13} \) Hz, respectively. Allowing $\gamma_{\rm min}$ to vary freely over the entire range
considered during model training (see Table \ref{tab:table_parameters}) results
in physically unrealistic values for other parameters, such as a very steep electron injection
spectrum. Therefore, we constrained $log(\gamma_{\rm min})$ to a narrow range between 1.5
and 2.0. During the fitting process, only the data
above $10^{12}$ Hz was considered, as emission in the radio band is
expected to be self-absorbed and should originate
from a different (extended) region of the jet. The modeling results are
presented on the left panel of Figure \ref{sed}, where the best-fit
model is indicated in red, while the
associated uncertainty in the model is depicted in gray.

The parameters derived from the fit are summarized in Table \ref{tab:param},
and the posterior distributions of these parameters are depicted in Appendix. The parameters we obtained are in agreement with those
typically estimated in FSRQ modeling. For instance, the power-law index of
the emitting electrons is $1.87$,
the
maximum energy of the injected electrons is $2.0 \times 10^3$, 
and the magnetic field strength is $0.20$ G. The Doppler boosting
factor, at $49.14$, is relatively high and not well constrained, reaching the
upper limit of our considered range. 
\citet{2021MNRAS.504.5074S} modeled the same period assuming an ad-hoc electron
distribution without considering cooling, and estimated a Doppler factor of
$50.8$. A high Doppler factor is necessary
to account for the pronounced Compton dominance observed in this
SED: the peak of the HE component surpasses the synchrotron/SSC components by
nearly two orders of magnitude.

We use the second neural network trained on the electron distribution function to display
the corresponding distribution on the bottom panel of Figure \ref{sed}. Clearly there is a large uncertainty on the electrons properties, in particular concerning its normalisation. In addition, it is clear that electrons do not cool below $\gamma_{\rm min}$, even though the external radiation field is quite intense. Since the maximum value of $\log \gamma_{\rm min}$ is bounded to 2 for this fit, this could potentially inhibit electron cooling at such small Lorentz factor, as it would require a strong magnetic field or a large density of the external photon field. In fact, the bottom left panel of Figure \ref{sed} shows the characteristic electrons Lorentz factors $\gamma_{\rm min}$ and $\gamma_{\rm max}$, alongside $\gamma_{\rm c,~synch} = 6\pi m_e c^2 / (B^2 R \sigma_{\rm T})$, representing the electron Lorentz factor at which electron would cool in a dynamical timescale by emitting synchrotron radiation. In this definition, $m_e$, $c$ and $\sigma_{\rm T}$ are the electron mass, the speed of light and the Thompson cross-section respectively. It is clear that this cooling effect is subdominant, as expected due to the dominance of the external Compton component.

\begin{figure*}
    \centering
    \includegraphics[width=0.48\textwidth]{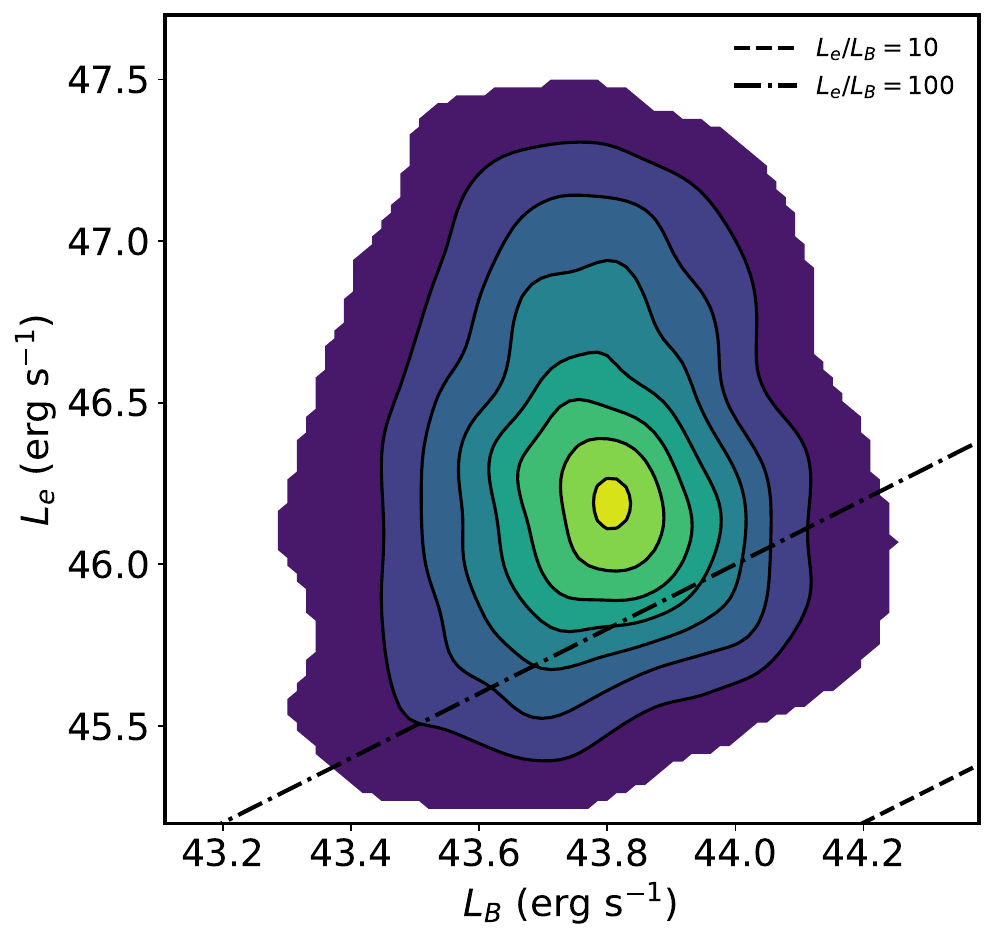}
     \includegraphics[width=0.48\textwidth]{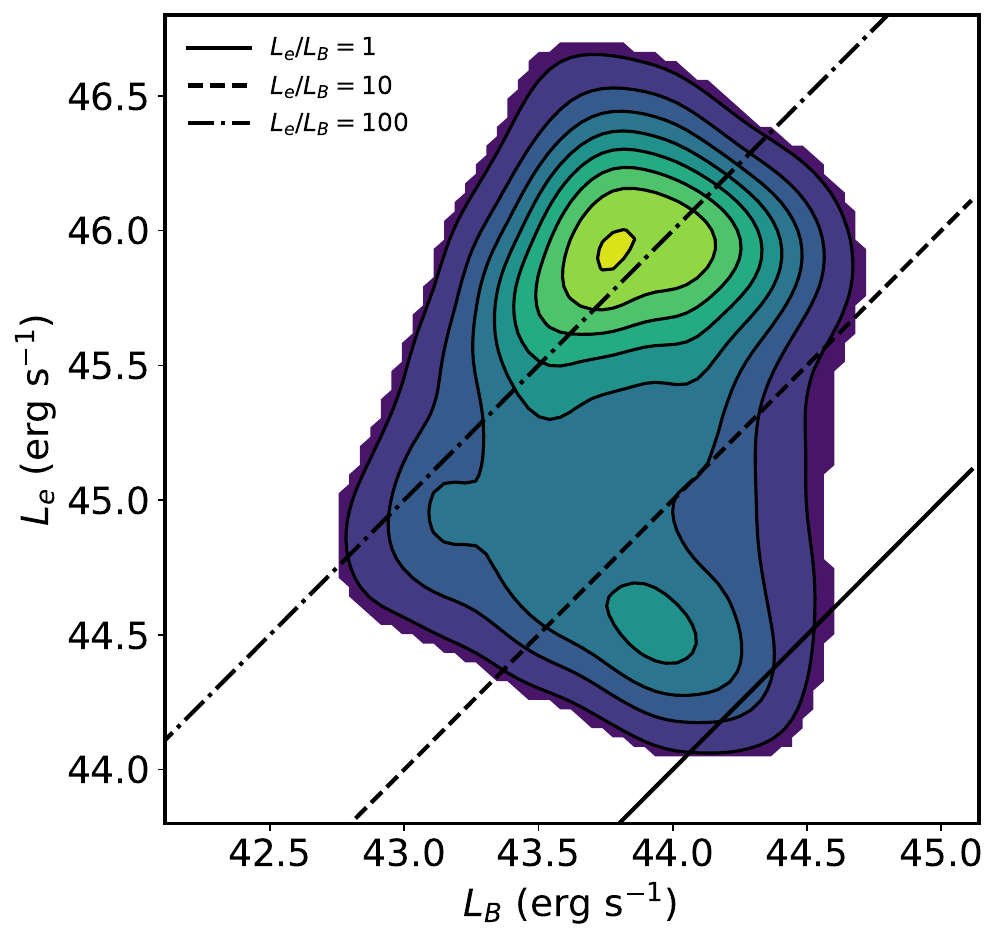}
    \caption{Comparison between the electron luminosity $L_e$ and the magnetic luminosity $L_B$ for 3C 454.3 (left) and CTA 102 (right) shown as the two dimensional projections of the posterior probability distributions. The emission of both sources originates in a plasma far from equipartition with $L_B \ll L_e$. Only CTA 102 seems to have a marginal region such that $L_B \lesssim Le$ in the bottom right. In fact, this region corresponds to the SSC scenario.}
    \label{Le_LB}
\end{figure*}

The fitting parameters also enable access to the energetic properties of the system. The disk luminosity,
estimated based on the requirements to explain the flaring GeV data, is
$L_{\rm d}=1.28\times10^{47}\:{\rm erg\:s^{-1}}$ (shown as dashed blue line in Figure \ref{sed} left panel), which is on the same order
as the Eddington luminosity for a black hole mass of
$M_{\rm BH}=1.5 \times 10^9 M_\odot$. However, it should be noted that
higher black hole masses for 3C 454.3 have also been estimated; for example,
\citet{2001MNRAS.327.1111G} estimated a black hole mass of $M_{\rm BH}=4 \times 10^9 M_\odot$.
This luminosity is only slightly higher, by a factor 2 or 3,
than the disc luminosity estimated based on the blue bump observed in the UV
band \citep[see e.g.,][]{2021MNRAS.504.5074S,2011MNRAS.410..368B}. The jet electron luminosity
is $L_{\rm e} = 5.61 \times 10^{46}\: \rm{erg\: s^{-1}}$, which is
$\sim 8\times10^2$ times higher than the magnetic field luminosity,
$L_{\rm B} = \pi c R^2 \delta^2 B^2 / 8\pi = 6.57 \times 10^{43}\: \rm{erg \: s^{-1}}$. This shows
that the system is far from equipartition. This is evident from the left panel of Figure
\ref{Le_LB}, where we show the two dimensional projections of the posterior probability distributions of $L_e$ and $L_b$. The dashed and dot-dashed lines represent $L_{\rm e}/L_{\rm B}=10$ and $L_{\rm e}/L_{\rm B}=100$, respectively. We note that the limits set on $\gamma_{\rm min}$ could influence the actual value of $L_e$ determined by the fit, and eventually change the equipartition ratio $L_e / L_B$, although we believe not substantially. We leave this detailed analysis for a future work. 

\subsection{CTA 102}

CTA 102, at a redshift of \(z = 1.037\) \citep{1965ApJ...141.1295S}, is
another prominent FSRQ that hosts a black hole with an
estimated mass of \(8.5 \times 10^8\: {\rm M_{BH}}\) 
\citep{2014Natur.510..126Z}.  The source periodically enters a state of active emission.
For instance, between 2016 and 2017, CTA 102 was in a prolonged active
state, exhibiting an extraordinary outburst across all wavebands
\citep[see Fig. 1 of][for the broadband evolution of the flux of
the source]{2022MNRAS.517.2757S}. On several occasions, the \gray\ flux
of the source exceeded \(10^{-5}\: {\rm photon\: cm^{-2}\: s^{-1}}\),
with emission episodes so bright that variability on a minute scale could be investigated
\citep{2018ApJ...854L..26S}. Furthermore, during these bright periods,
the spectrum of the source deviates from a simple power-law model
\citep{2020A&A...635A..25S}, allowing to investigate the location
of the emission region and/or the cutoff in particle emissions.
The emissions from CTA 102 have been examined considering various
models, including among others different locations of the emission region
\citep[within or outside the BLR, 
e.g.,][]{2018ApJ...863..114G, 2018ApJ...866...16P, 2020A&A...635A..25S},
in the context of the twisted inhomogeneous jet model
\citep{2017Natur.552..374R}, or with a superluminal
component crossing a recollimation shock \citep{2019A&A...622A.158C}.

The SED of CTA 102 between MJD 56196.7-56202.3 was retrieved from \citet{2022MNRAS.517.2757S}. It is displayed alongside with the results of the modeling on the right panel of Figure \ref{sed}.  
Contrary to the the previous model, we set the minimum Lorentz factor to $\gamma_{\rm min}=100$ since no
low energy data are available to help constraining its value. In addition, we set
the temperatures of the external photon fields
to \( \nu_{\rm BLR} = 2.47 \times 10^{15} \) Hz, and
\( \nu_{\rm IR} = 3 \times 10^{13} \) Hz. Moreover, we impose \({\rm M_{BH}}=8.5 \times 10^8\: {\rm M_\odot}\). The results of the modeling are provided in Table
\ref{params}, and the model with the best parameters, i.e., where the
likelihood is maximum, is highlighted in red  on the right panel of
Figure \ref{sed}. From the red curve, it is clear that the combined contribution of all components effectively
explains the observed data. The power-law index of injected electrons, $p=1.97$, is mostly determined from the SSC contribution to the X-ray band while the maximum electron Lorentz factor $\gamma_{\rm max}=1.52\times10^3$ is constrained by data in the optical/UV and GeV band. In this case, the estimated size of the emitting region, and consequently its distance from the central object, is $4.68\times10^{15}$ cm. In addition, the disk luminosity is $6.25\times10^{45}\:{\rm erg\:s^{-1}}$, which is slightly lower than $L_{\rm d}=4.14\times10^{46}\:{\rm erg\:s^{-1}}$, inferred from the luminosity of the Ly$\alpha$ line \citep{2005MNRAS.361..919P}. The comparison of jet luminosity in electrons and in the magnetic field is shown in the right panel of Figure \ref{Le_LB}, which  demonstrates that, similarly to 3C 454.3, the emitting region is far from equipartition.

As the jet propagates through the external photon fields characterized by a high comoving energy density, the injected electrons are interacting with these photons, producing the HE component. We display the uncertainty on the electron distribution function at equilibrium in the right bottom panel of Figure \ref{sed}, alongside the results from the best fit model in red. From this figure, it is seen that electrons are in the slow cooling regime for our choice of $\gamma_{\rm min}$, which could not be constrained independently because of the lack of data between the optical band and the radio (not associated to the jet emission).

\begin{figure}
    \centering
    \begin{tabular}{cc} 
    \includegraphics[width=0.48\textwidth]{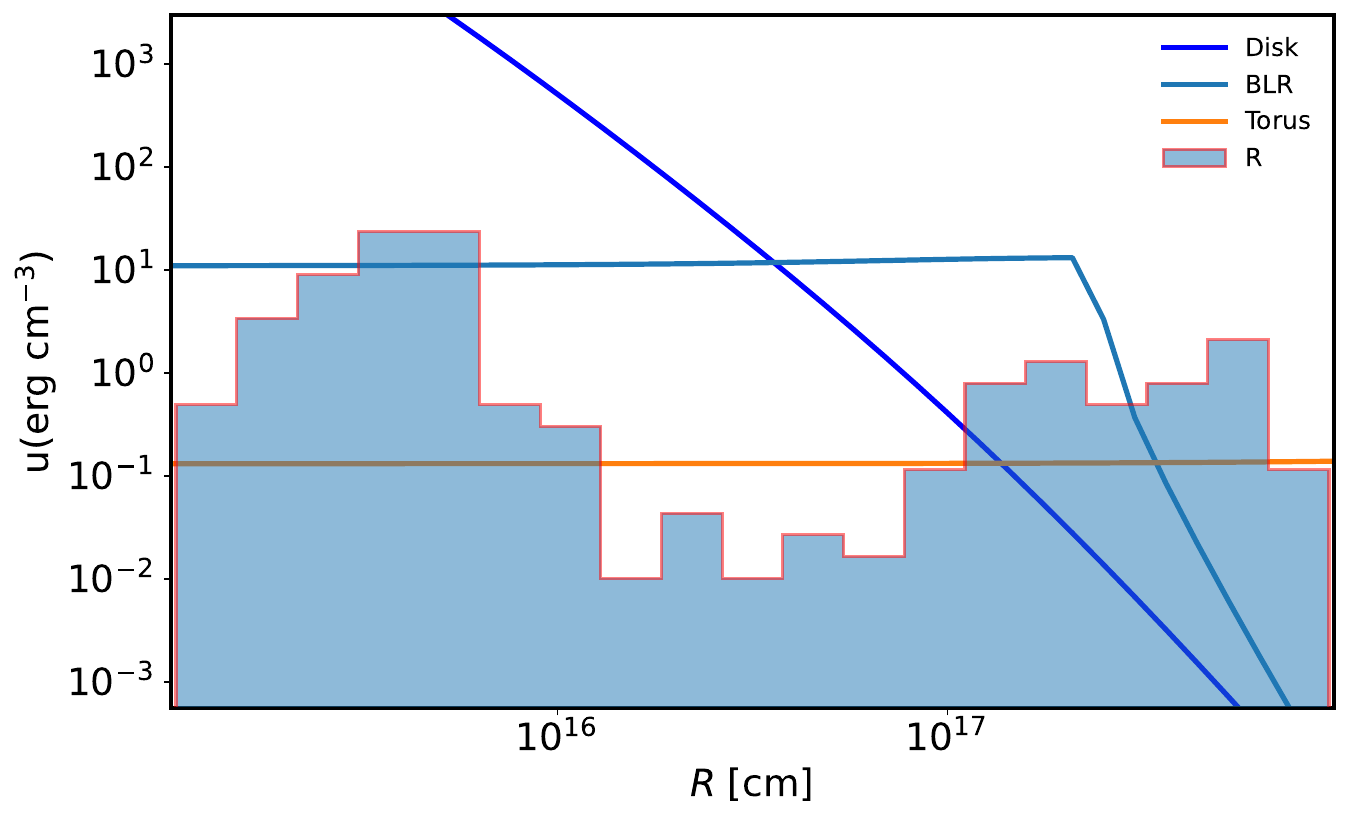}
    \end{tabular}
    \caption{Variation of the external photon field energy density with the radius of the emission region. Contributions from the disk, the BLR, and the torus are represented in blue, orange, and green, respectively. The posterior distribution of the radius $R$ is shown as the light blue histogram, displaying the bi-modality of the posterior. The emission can either be produced at small radius where the disk internal energy density dominates, or at large radius such that all external fields components are small, and the emission is purely from an SSC origin.}
    \label{fig:elec}
\end{figure}

%%%%%%%%%%%%%%%%%%%%%%%
The posterior distributions of the model
parameters shown in Appendix, reveal a bimodal distribution for the radius $R$, at  $R\sim 4\times10^{15}$ cm and $R \sim 5\times10^{17}$ cm.
The location of the emission region, and consequently the main field
contributing to the formation of the HE emission in FSRQs, remains an open
question. In our framework, the fit drives the selection of the
relevant photon field to explain the provided data by
finding suitable parameter sets.  This
bimodality suggests the existence of two sets of conditions compatible with modeling
the observed SED of CTA 102. Specifically, this indicates two distinct
physical scenarios with different geometrical locations within the blazar jet that can produce the emission. In Figure \ref{fig:elec}, the comoving energy densities of all external photon fields considered in our model are displayed as a function of radius for a set Lorentz factor, alongside the posterior distribution of the radius obtained from our fit. We note that these calculations were performed for $\delta=40.74$. Changes in $\delta$ affects all external fields nearly equally, meaning that the curves in Figure \ref{fig:elec} will only scale up or down, while the relationships among all fields will remain unchanged. For the first
maximum of the radius posterior distribution at $R\sim 4\times10^{15}$ cm, which corresponding model is shown in red in the right panel of Figure \ref{sed},  the photon comoving energy density is dominated by photons from the disk, indicating that the HE is due to external Compton. However, the other maximum indicates a significantly larger radius ($\sim5\times10^{17}$ cm). This
parameter set is also characterized by a  higher $L_{\rm e}$ and $\gamma_{\rm max}$ but a
lower Doppler boost $\delta$. For this scenario, the contributions of external
fields are reduced due to the much larger radius, the electron energy density is higher while
the magnetic energy density is lower. Therefore, the emission in the HE band is attributed
to the SSC emission for this parameter set. We show the most probable spectrum from this parameter mode by the red dashed line in the right panel of Figure \ref{sed}.

Our method illustrates the complexity of modeling the SED
of FSRQs and highlights the advantages of our applied model. Unlike the commonly
accepted approach that presupposes the contributing field to the formation of HE
emission, our model allows the emission location to remain unspecified. It considers
all photon fields, identifying the best possibility for explaining
the data. Our approach also presents the flexibility of setting the disk luminosity $L_d$ and the radius $R$ such as to select beforehand the dominant external field component to be used by the model and obtain the relevant parameter distributions under these assumptions.

\section{Conclusion}\label{sec:conc}

In this data-rich era of blazar research, the modeling of their SEDs
using numerically intensive models is becoming essential. This approach
is the principle way to thoroughly investigate the
processes occurring within their jets. In \citet{BSD23}, we introduced
a novel framework for fitting the SED of blazars. This framework uses a
CNN trained on a large set of SSC spectra generated by \textit{SOPRANO}
which accurately reproduces the radiative signatures of the model and
is used to fit observational data and infer model parameters. Here, we
extend this framework by training the CNN on a more complex model that
also includes external photon fields for inverse Compton scattering and pair creation.  

In this study, the CNN is trained on spectra generated by \textit{SOPRANO},
incorporating the most relevant external photon fields (direct disk photons,
photons from the BLR, and the dusty torus) and internal photon fields
(synchrotron photons), as well as all relevant cooling processes and the
pair creation process. We have shown that, despite increased complexity
(7 versus 11 free parameters for the SSC and EIC models, respectively),
slightly modifying the layers and dimensions of the CNN enables
accurate reproduction of the model spectral features. The CNN
execution time is significantly shorter compared to computations with
\textit{SOPRANO} (or any other code that self-consistently treats particle
acceleration and injection), allowing it to be effectively coupled with
fitting techniques to derive parameters that best describe the data. The
application of the CNN has been demonstrated through fitting the broadband
SEDs of the well-known FSRQs 3C 454.3 and CTA 102 during flaring
periods, thereby constraining the parameters of the EIC model and
obtaining their posterior distributions.

Although the model satisfactorily reproduces the radiative signature of
particle emissions when the emission region is within an intense radiation
field, several limitations had to be considered. Firstly,
the model has been trained for a wide range of parameters, which are close
to physically realistic scenarios. Specifically, the disk temperature,
and consequently the energy of disk photons, is determined from the black hole
mass and accretion disk luminosity, both entailed with large uncertainties. Additionally, the temperatures of the BLR and the torus are free parameters that can vary within a certain range.
Treating these parameters as free variables in the fitting process allows
for a wider selection range, which aids in converging towards the optimal
fit parameters. Conversely, fixing these parameters
—particularly the black hole mass— to their estimated
observational values could lead to more stringent constraints on the
remaining free parameters.

Our model also suffers from a limitation: by
linking the disk temperature to the black hole mass
and accretion disk luminosity imposes limitations on accurately
reproducing emissions from a hot accretion disk. \citet{GTG09} suggested
that the blue bump observed in the UV band could be attributed to
thermal disk emission. To achieve a disk component peaking at $\sim10^{15}$ Hz
under the disk model considered here,
a luminosity $L_{\rm d}\geq10^{48} , \rm{erg , s^{-1}}$ would be
required. This would result in a very high flux of the disk component with respect to the jet emission. This limitation boils down to assuming that the radiative efficiency of the disk $\eta$ is set to $\eta = 0.1$. However, keeping this limitation in mind, the approach can be
successfully applied to high-redshift blazars
\citep[see, e.g.,][]{GTG09, 2020MNRAS.498.2594S, 2024MNRAS.tmp..315S}. We plan to train a
new CNN for which the disk emission will
be modeled as a simple black body with  independent $L_{\rm d}$
and temperature to accurately model cases where an excess in the UV band is observed, at the expense of losing a straightforward physical interpretation.

The CNN presented here, together with the one from \citet{BSD23}, constitutes
a comprehensive and flexible methodology for self-consistent blazar modeling
within leptonic scenarios. These models collectively enable in-depth
investigations of the emission from both BL Lac and FSRQs,
and hold the potential to advance our
understanding of the physics of relativistic jets. Similar to the SSC model,
the EIC model will be made publicly available through the MMDC. This
platform permits researchers to upload their data and perform self-consistent
modeling, with both the SSC and EIC
models, through a dedicated web interface, providing them with the
best-fitting parameters and their posterior distributions.

In summary,  in this study, we demonstrated that
the methodology presented in \citet{BSD23} can be extended and applied to
more complex physical scenarios, which involve more free parameters
and require taking into account the contribution of several photon fields.
As a natural next step, this methodology will be expanded to include
the even more sophisticated hadronic, and lepto-hadronic models.
In these models, along with the contribution of electrons, the contributions
of protons and secondaries produced in photo-pion and photo-pair interactions
need to be considered. Training a CNN on these new models requires careful
treatment of the interactions of all primary and secondary particles. This
is not a trivial task but is essential for interpreting multi-messenger data
from blazar observations, including both photons and neutrinos.

\section*{Acknowledgements}
NS, SG and MK acknowledge the support by the Higher Education and Science
Committee of the Republic of Armenia, in the frames of the
research project No 23LCG-1C004.
DB, HDB and AP acknowledge support from the European Research Council via the
ERC consolidating grant $\sharp$773062 (acronym O.M.J.). We acknowledge with gratitude that Altair has agreed to pay for the generation of the training sample on AWS EC2.

\section*{Data availability}
All the observational data used in this paper is public. The convolutional neural network used
to fit the SEDs can be shared on a reasonable request to the corresponding author. In addition, it is publicly available through the Markarian Multiwavelength
Datacenter (\url{http://www.mmdc.am}).

\bibliographystyle{mnras}
\bibliography{biblio}

\appendix

\section{Parameter posterior for 3C 454.3 and CTA 102}

\begin{figure*}
    \centering
    \includegraphics[width=0.95\textwidth]{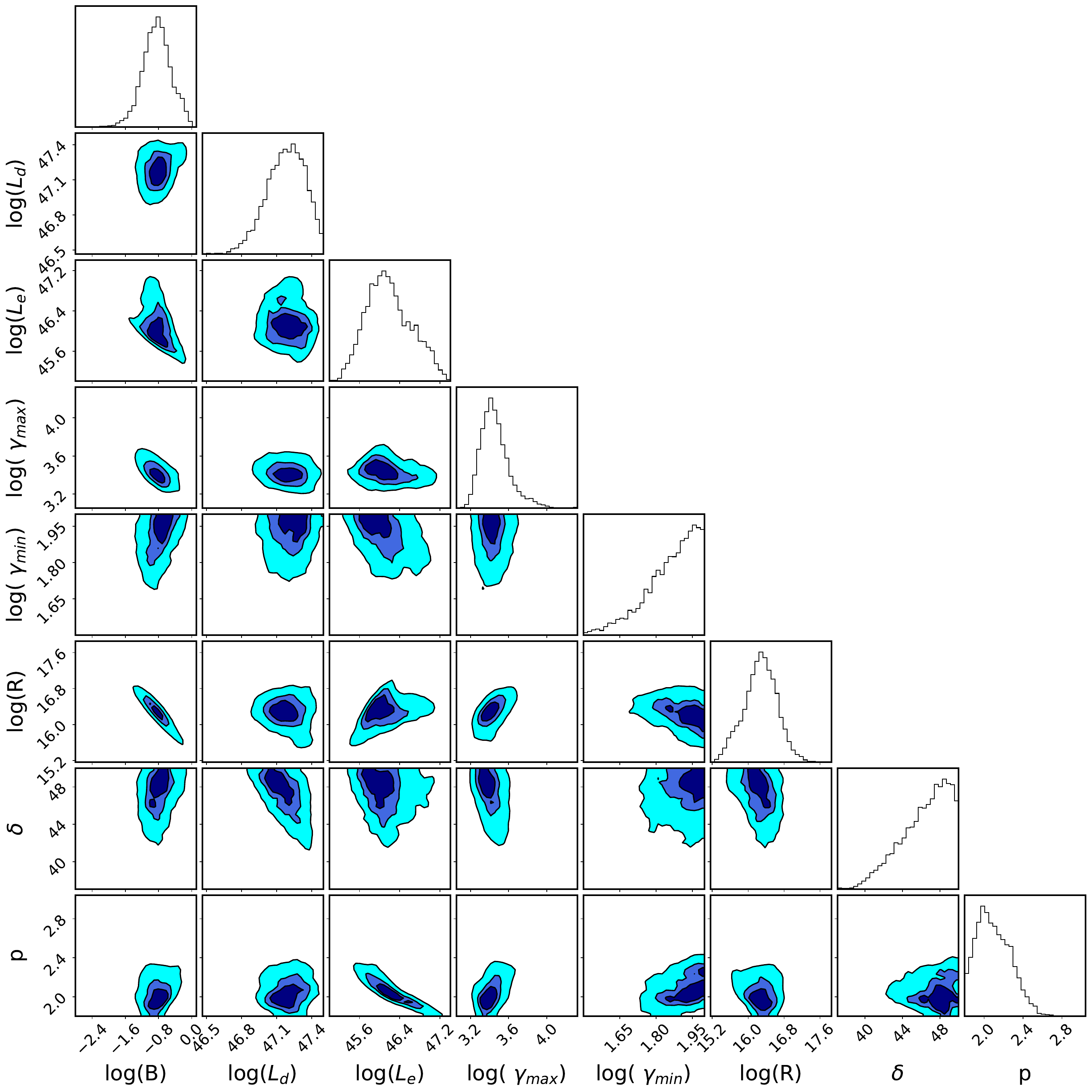}
    \caption{ Parameter posterior distributions from the SED modeling of 3C 454.3 for the period MJD 55519.59-55520.19. With the exception of the Doppler boost $\delta$ which get close to the parameter boundary, all parameters are well-constrained. }
    \label{fig:posterior_3C454}
\end{figure*}

\begin{figure*}
    \centering
    \includegraphics[width=0.95\textwidth]{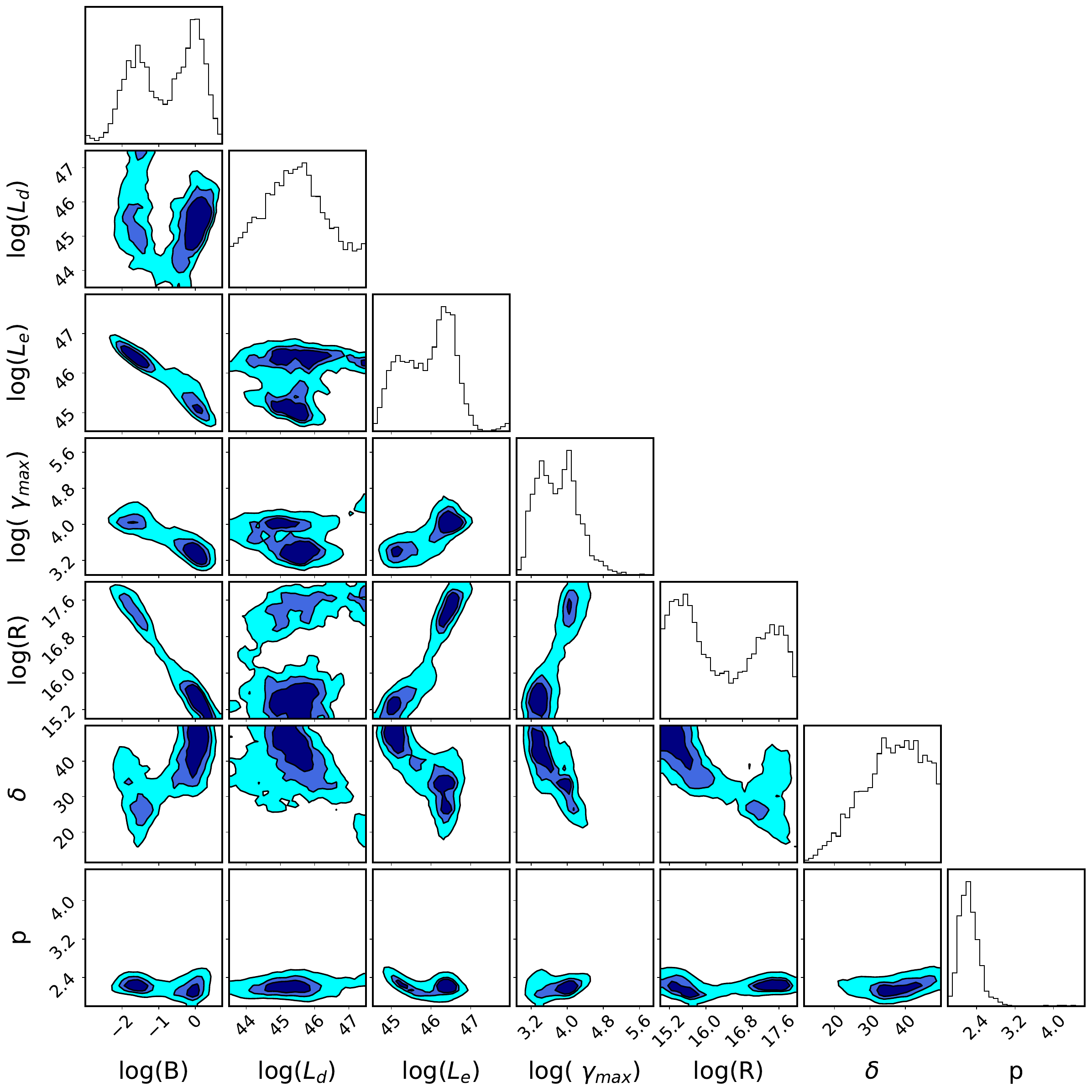}
    \caption{Parameter posterior distributions for CTA 102, showing a bimodal distribution for the radius $R$ alongside other parameters. For this fit, $\gamma_{\rm min}$ was set to $10^2$.}
    \label{fig:posterior_cta102}
\end{figure*}

We show in this appendix the parameter posterior distributions of 3C 454.3 and CTA 102 in Figure \ref{fig:posterior_3C454} and \ref{fig:posterior_cta102} respectively.

\end{document}